# Using Ecological Propensity Score to Adjust for Missing Confounders in Small Area Studies

Yingbo Wang
*Novartis, Basel (Switzerland)*
Sylvia Richardson
*MRC Biostatistics Unit, Cambridge University (UK)*
Monica Pirani, Anna Hansell, Marta Blangiardo*
*MRC-PHE Centre for Environment and Health, Imperial College London (UK)*
m.blangiardo@imperial.ac.uk
∗ To whom correspondence should be addressed.

January 18, 2017

## Abstract

Small area ecological studies are commonly used in epidemiology to assess the impact of area level risk factors on health outcomes when data are only available in an aggregated form. However the resulting estimates are often biased due to unmeasured confounders, which typically are not available from the standard administrative registries used for these studies. Extra information on confounders can be provided through external datasets such as surveys or cohorts, where the data are available at the individual level rather than at the area level; however such data typically lack the geographical coverage of administrative registries. We develop a framework of analysis which combines ecological and individual level data from different sources to provide an adjusted estimate of area level risk factors which is less biased. Our method (i) summarises all available individual level confounders into an area level scalar variable, which we call *ecological propensity score* (EPS), (ii) implements a hierarchical structured approach to predict the values of EPS whenever they are missing, (iii) includes the estimated and predicted EPS into the ecological regression linking the risk factors to the health outcome. Through a simulation study we show that integrating individual level data into small area analyses via EPS is a promising method to reduce the bias intrinsic in ecological studies due to unmeasured confounders; we also apply the method to a real case study to evaluate the effect of air pollution on coronary heart disease hospital admissions in Greater London. Hierarchical model, observational study, spatial statistics, missing data, environmental epidemiology.

# 1 Introduction

Small area studies are commonly used in epidemiology to investigate the spatial variation of a health condition across the population or to evaluate the geographic patterns of diseases in relation to environmental, demographic and socio-economic factors.

These studies are based on administrative registries, which are characterised by good spatial coverage for large populations, but usually only record a very limited set of information (typically age, sex and spatial location) and thus miss important confounders (e.g. data on lung cancer or respiratory diseases from Hospital Episode Statistics databases do not include information about smoking), potentially leading to biased estimates of the effects of risk factors (e.g. air pollution).

In this paper, we integrate data from administrative registries with cohorts / surveys, which contain detailed information on participants. Through this, the inferences drawn at the ecological level will take advantage of the population representativeness and of the statistical power from the administrative registries, but at the same time will allow to adjust the effect estimates for all the potential confounders available from the cohorts / surveys.

Registries contain data on each individual in the target population, while cohorts / surveys typically cover only a subset of individuals; for this reason confounders obtained from the latter source will be partially measured and available only on a subset of areas, leading to a missing data issue which needs to be tackled. Multiple imputation (MI), pioneered by Rubin (1987) is probably the most common strategy to deal with this issue and consists of a Monte Carlo simulation to replace the missing values with a relatively small number of simulated versions. The two main approaches to implement MI are joint modelling and chained equations. As Hughes et al. (2014) showed that the difference in the results from the two approaches are negligible, chained equations are preferred as they are based on conditional distributions, enabling fast computations. As an example, the Multiple Imputation using Chained Equation (MICE) proposed by Buuren and Oudshoorn (1999) fixes initial values for all the missing variables and regress each of these against the remaining ones cyclically a number of times (see White et al. 2011 for a thorough review of the method). Then, in a separate step, the imputed variables are included in the substantive model, which evaluates the link between the exposure / risk factor and health outcome. Such approach can suffer from lack of "congeniality": Meng (1994) states that for a model to be congenial the imputation model needs to include the same variables (including the dependent variable) as those of the substantive model, in order to avoid estimates biased toward the null; this might be non trivial for non linear relationships between the outcome and the exposure / risk factors, which is the case when the outcome is available in the form of aggregated counts as in small area studies. Furthermore, such method is not suited to add spatial dependency in the imputation model, which may lead to bias if the missing values are geographically related, as it is often the case in a small area study.

In a Bayesian framework, missing data imputation is generally considered via the integration of the imputation and the analysis models in a coherent global analysis (see for instance Molitor et al. 2009, Daniels and Hogan 2008). This approach benefits from extreme flexibility, but entails heavy computational burden if there are more than a few missing covariates, de facto requiring oversimplifications of the epidemiological problem, as in reality the number of potential partially measured confounders is typically large.

To avoid high dimensional imputation, a viable alternative summarises the partially measured confounders from the cohorts / surveys through a composite index, so that only one variable needs to be imputed. In this perspective, the propensity score (PS) has been suggested, in the form of a calibration model proposed by Stürmer et al. (2005) as well as in a Bayesian framework developed by McCandless et al. (2012). The first approach estimates a gold standard PS on the units with information on the partially measured confounders, while an "error-prone" PS is estimated on all the units. Then Stürmer et al. (2005) specify a regression model to estimate the relationship between the gold standard PS and the error prone PS in order to predict the gold standard where missing. Note that the proposed calibration model assumes a linear relationship between the gold standard PS and the error prone PS; moreover the outcome variables is not included in the regression (in contrast to what is recommended in the missing data literature). The second method is a Bayesian Propensity Score (BayesPS) formulated to include information from a cohort with fully observed confounders in an individual level study based on administrative data. A propensity score is built on the cohort data to summarise the confounders. On the individuals who do not have information on the confounders, the approach then imputes the propensity score from its empirical distribution. This strategy can be applied effectively regardless of the number of unmeasured confounders.

In this paper, we develop a novel Bayesian modelling framework to fit the propensity score on small area studies and ecological covariates. In particular (i) we propose an imputation model for areas with a missing propensity score; this accounts for the spatial structure of the data and can easily accomodate non-linearity in the relationship with other variables; (ii) we include the estimated / imputed propensity score in the analysis model in a flexible way to provide effective confounder adjustment when evaluating the effect of a risk factor on a health outcome; (iii) we discuss and account for the different sources of feedback across the overall modelling framework. The structure of this paper is as follows: Section 2 introduces our proposed *ecological propensity score* (EPS) framework for small area studies and Section 3 presents an extensive simulation study to evaluate the performance of the developed approach and to compare it with the commonly used MICE for imputing missing data; Section 4 applies this method to assess the link between airborne particle pollution and coronary heart disease hospital admissions in Greater London; finally we include a discussion and concluding remarks.

## 2 The ecological propensity score (EPS)

Before starting with the description of the EPS framework we set the notation used throughout the paper: a) $i \in S$ identifies the set of areas with coverage from the survey or cohort; b) $i \in I$ identifies the set of areas without coverage from the survey or cohort; c) $O_i$, $E_i$ are the number of health outcome occurrences, which are observed and expected (using standardised rates); they are available on all the areas ($i \in I \cup S$); d) $X_i$ is the exposure or risk factor of interest observed on all the areas ($i \in I \cup S$); e) $\boldsymbol{C}_i$ is the set of P area level confounders observed on all the areas ($i \in I \cup S$); f) $\boldsymbol{M}_i$ is the set of Q area level confounders which are unobserved on all the areas ($i \in I \cup S$), but for which individual level data $\boldsymbol{m}_{qi} = \{m_{qi1}, \ldots, m_{qin}\}$ are available on $i \in S$ from the survey or cohort.

To develop the modelling framework we consider three steps: a) **EPS estimation** - in the areas with available individual level information on the full set of confounders ($i \in S$) the information is aggregated through small area estimation models and the ecological propensity score is estimated using these ecological confounders. *b)* **EPS imputation** - in the areas without survey coverage ($i \in I$) a Bayesian imputation model is set up to predict EPS based (i) on the relationship between the estimated EPS and other variables in the $S$ survey areas and (ii) on the spatial structure of the data. *c)* **EPS adjustment** - the EPS (estimated or imputed) for all the areas is then included in the analysis model providing confounder adjustment to the ecological relationship of interest between $X$ and $O$.

As our approach is formulated in a Bayesian framework a joint model could be specified for the three components; however when using propensity score approaches to adjust for confounder effects, may be necessary to prevent feedback in estimating some parameters to avoid biases (Zigler et al., 2013). In the remainder of this section we describe the model in details and discuss the issues around feedback.

## 2.1 EPS estimate

The first step of the EPS framework uses the individual level confounding information to estimate corresponding (latent) area level confounders on the survey areas; as the survey samples are typically small and at the same time they are often spatially distributed, we carry out this estimation by using a hierarchical spatial model for all the confounders allowing for each survey area to be embedded in a larger spatial unit $d_i$, e.g. used for the survey stratification, to ensure complete spatial coverage. From the survey (or cohort), $n$ values $\{m_{qi1}, \ldots, m_{qin}\}$ are observed on the $i$-th area ($i \in S$), for the $q$-th confounder variable; such variables can be dichotomous (e.g. smoking status), categorical (e.g. education level) or continuous (e.g. BMI) so that it is reasonable to assume that $m_{qi1}$ belongs to the exponential family (e.g. following a Binomial, Poisson, Normal distribution); hence to estimate $M_{qi}$ we specify the link function as follows:

$$g(M_{qi}) = \alpha_q + \psi_{1qd_i} + \psi_{2qd_i} + v_{qi} \tag{1}$$

where $\boldsymbol{\alpha} = (\alpha_1, \ldots, \alpha_Q)$ is the intercept for each confounder and $\boldsymbol{\psi}_{qd_i}$ model the spatially structured and unstructured variability for the larger spatial units ($d = 1, \ldots, D$); a multivariate version of the BYM prior (MVBYM) proposed by Besag et al. (1991) is specified to account for spatial dependency for each confounders as well as for correlation between these (see Appendix A for details of MVBYM prior). In addition, $v_{qi} \sim \text{Normal}(0, \omega_v^2)$ captures the residual overdispersion at the inference unit level.

EPS is derived following the specification of McCandless et al. (2012) :

$$\begin{aligned}
\text{logit}(P(X_i = 1 | \mathbf{C}_i, \boldsymbol{M}_i)) &= \theta_1 + \mathbf{C}_i^T \boldsymbol{\theta}_C + \boldsymbol{M}_i^T \boldsymbol{\theta}_M \\
\text{EPS}_i &= \boldsymbol{M}_i^T \boldsymbol{\theta}_M
\end{aligned} \tag{2}$$

where $X$ needs to be dichotomised and $\boldsymbol{\theta}_C = (\theta_{C_1}, \ldots, \theta_{C_P})$, $\boldsymbol{\theta}_M = (\theta_{M_1}, \ldots, \theta_{M_Q})$, while $\boldsymbol{M}_i$ are the area level estimates for the Q confounders obtained in (1).

Here feedback from $X$ to $\boldsymbol{M}$ in (2) needs to be cut as it would distort the estimates of $\boldsymbol{M}$ in (1). Nevertheless uncertainty on $\boldsymbol{M}$ is fed forward from (1) to the EPS estimate in (2), which thus has a mixing distribution over the uncertainty in $\boldsymbol{M}$. A summary of that mixing distribution will then be used in the imputation model as detailed below.

## 2.2 EPS imputation

The EPS estimation presented in (2) can be used only on the survey areas; when incomplete areas are present in the study region an additional step is needed to impute EPS due to the fact that the individual level covariates and consequently $M_i$ are missing. Thus for each area $i$ ($i \in S \cup I$) we specify:

$$\text{EPS}_i \sim \text{Normal}(\ \eta_1 + f(\mathbf{C}_i) + \gamma X_i + \phi_{d_i},\ \sigma^2_{EPS}). \tag{3}$$

where $\text{EPS}_i$ is the posterior mean from the estimation in (2) for $i \in S$ and is missing for $i \in I$. In (3), the link between $X$ and $M$ as written in (2) is reversed and now specifies a relationship between the posterior mean for EPS and (i) the dichotomous $X$, (ii) the confounders $\mathbf{C}$ through the function $f(.)$. This function should meet three criteria: *a)* be data-driven, as (3) does not have an epidemiological interpretation, *b)* be able to cope with a high dimensional $\mathbf{C} = (C_1, \ldots, C_P)$, *c)* be easy to scale up to over thousands of observational units if needed. In order to satisfy these three requirements, we choose a second order random walk (RW(2)) as functional form for $f()$ to link each $\mathbf{C}$ to (3), which was initially proposed by Fahrmeir and Lang (2001) to adjust for non-linearity.

An additional random effect $\phi$ is included to account for residual variation using the same spatial resolution as in (1) to ensure full spatial coverage. As it is likely that these residuals will exhibit a spatial structure, a conditional autoregressive model is specified on $\phi$ based on neighborhood similarities, through the univariate formulation of the BYM included in (1).

The $\text{EPS}_i$ for $i \in I$ is predicted using the relationship with $\mathbf{C}$, $X$ and the spatial structure estimated on $i \in S$.

## 2.3 EPS adjustment

The analysis model is specified in a small area framework to evaluate the direct (area level) effect of a risk factor (or exposure) $X$ on a health end point $O$ after adjustment for observed confounders $C$ and unobserved confounders $M$ using EPS. In its typical formulation, for each area ($i \in S \cup I$) the number of cases of the health outcome $O_i$ is modelled as follows:

$$\begin{aligned} O_i &\sim \text{Poisson}(E_i \lambda_i) \\ \log(\lambda_i) &= \beta_1 + \beta_2 X_i + \mathbf{C}_i^T \boldsymbol{\beta}_C + h(\text{EPS}_i) + \xi_{1i} + \xi_{2i} \end{aligned} \tag{4}$$

where $E_i$ represents the expected number of cases obtained from standardised rates. The parameter $\xi_{1i}$ accounts for over-dispersion, while $\xi_{2i}$ accounts for spatially structured variation and is typically modelled through a conditional autoregressive structure similarly to (1) and (3).

The $\text{EPS}_i$ feeds into (4) as:

- $\boldsymbol{M}_i \boldsymbol{\theta}_M$ estimated for $i \in S$ from (2);

- the posterior predictive distribution from (3) for $i \in I$.

To be able to estimate a direct effect of $X$ on $O$ we want to make sure that EPS is included in the analysis model through a non-linear flexible function $h(\cdot)$, which would guarantee the best approximation for its relationship with $O$; similarly to the imputation model specification, we assign a RW(2) to link the EPS into the analysis model. Note that (3) and (4) are jointly estimated, so that the outcome is allowed to influence the missing EPS imputation via its feedback, as recommended in the missing data literature (see for instance Kenward and Carpenter 2007); at the same time uncertainty from the EPS imputation is carried forward into the analysis model. A graphical representation of the EPS framework including where feedback is allowed/cut is visualised in Figure 1.

[Figure 1 here]

## 2.4 Specifying priors

To complete the model specification, prior distributions need to be assigned. For MVBYM in Model 1, the priors for $\boldsymbol{\Sigma}_{\psi_1}^{-1}$ or $\boldsymbol{\Sigma}_{\psi_2}^{-1}$ are chosen to be Wishart$(\nu, \mathbf{W})$ centred around the empirical variance $\hat{\boldsymbol{\Sigma}}$ by specifying $\mathbf{W}_{\psi_1} = \nu_{\psi_1}\hat{\boldsymbol{\Sigma}}$ or $\mathbf{W}_{\psi_2} = \nu_{\psi_2}\hat{\boldsymbol{\Sigma}}$. In general, an empirical variance $\hat{\boldsymbol{\Sigma}}$ is considered to be an approximate estimator of the true variance $E(\boldsymbol{\Sigma})$. This approximation is acceptable because the degree of freedom $\nu$ is typically chosen to be the minimum to ensure that the specified Wishart is a diffused prior, i.e. $\nu_{\psi_1} = \nu_{\psi_2} = Q$ where Q is the number of confounders as presented in Section 2.1 (Lunn et al. 2012).

The priors for the coefficients of the logistic regression in step 1 are chosen to cover the odds increase / decrease within 15 fold, which is a reasonable assumption in epidemiological studies (Greenland 2005), for instance $\theta_{M_q}, \theta_{C_p} \sim N(0, \frac{log(15)}{2})$. The rest of priors are chosen to be minimally informative, i.e. the regression coefficients are modelled as Normal$(0, 10^3)$ and the standard deviation of the random effects are modelled as Uniform$(0, 1000)$.

All the models were implemented in the BUGS language (Lunn et al. 2012).

## 3 Simulation

In this section, we present a simulation study to evaluate the performance of the EPS framework and to compare it with the commonly used MICE approach (Buuren and Oudshoorn 1999), which we implemented through the corresponding miceR package.

We consider three scenarios: *1)* the missing confounders $\boldsymbol{M}_i = (M_{1i}, \ldots, M_{Qi})$ are available in all areas ($i \in S \cup I$), and this is the benchmark (Scenario 1); *2)* the missing confounders $\boldsymbol{M}_i$ are assumed to be available in some areas ($i \in S$, Scenario 2); *3)* the missing confounders $\boldsymbol{M}_i$ are not directly available, and these confounders might be estimated through individual level data from surveys / cohorts (only on the survey areas, $i \in S$, Scenario 3). To assess the impact of the sample size of individual level data on the exposure estimation, Scenario 3 includes different numbers of individuals (5, 10, 20, 100) sampled from the survey / cohort in each area.

The variables are simulated in the following sequence on 300 areas: (i) two confounders $\mathbf{C}$ and four confounders $\boldsymbol{M}$ are generated from the inverse logit transformation of multivariate normal distributions, (ii) $X$ is simulated based on $\mathbf{C}$ and $\boldsymbol{M}$ (using (2)), (iii) $O$ is simulated from a Poisson with $E = 100$ (to mimic the real case we are illustrating in the next section), $X$ and confounders $\mathbf{C}$ and $\boldsymbol{M}$, (iv) for each area $n$ individual data $\boldsymbol{m}_{qi} = (m_{qi1}, \ldots, m_{qin})$ are simulated from a Bernoulli distribution based on the proportion $M_{qi}$. To mimic the limited survey coverage, Missing At Random (MAR) criterion is applied to remove $\boldsymbol{M}$ (for scenario 2) or $\boldsymbol{m}$ (for scenario 3) from around 50% of the areas. The detailed simulation process is described in Appendix B. This process is repeated 100 times to create 100 paired ecological level and individual level datasets.

As MICE cannot account for spatial dependency, we did not include spatial random effects in our simulation; additionally at present MICE cannot be linked to a multilevel model to estimate $\boldsymbol{M}$ as in (1), thus we extract the posterior means of $\boldsymbol{M}$ provided by (1), and then plug these into MICE. The outcome is included into the imputation model by adding the Standardised Mortality Ratio (i.e. SMR= $O/E$) as an extra predictor. The missing $\boldsymbol{M}$ are imputed ten times and included in the analysis model to estimate the exposure effect parameter $\beta_2$ using the same formulation as (4).

Then the ten estimated $\beta_2$ are summarised through Rubin's combination rule (Rubin 1987).

Bias, root mean squared error and width of the 95% credibility interval (CI95%) are used to compare the performance of the simulation scenarios; the true $\beta_2$ is chosen to be 0.5, which corresponds to a 64.9% ((exp(0.5)−1)×100) increment in the health risk for high vs low exposure.

## 3.1 Results

Table 1 presents the results of the simulation study when the true value of $\beta_2 = 0.5$. The benchmark (scenario 1) assumes the availability of confounders $\boldsymbol{M}$ in all areas, which allows to evaluate the impact of using EPS as a summary index instead of including each confounder separately into the analysis model. The estimation from EPS adjustment is almost identical to the regression approach which directly includes $\boldsymbol{M}$ as covariates. The CI95% width of $\beta_2$ from the EPS adjustment model is slightly wider than that of the regression adjustment, and this agrees with Senn et al. (2007) which showed that the estimation from PS stratification (very close to PS adjustment) is more conservative than that from the direct regression analysis.

Ignoring the information from the confounders $\boldsymbol{M}$ allows us to evaluate what size of bias we are potentially dealing with (naïve case). Since $\boldsymbol{M}$ is simulated to be a confounder for the relationship between outcome and exposure, ignoring it leads to a serious bias and RMSE in estimating $\beta_2$ (both equal to 0.28), as well as to a small uncertainty around it (width of CI95% equal to 0.049, which is smaller than in any other case).

Scenario 2 and 3 assume the partial availability of $\boldsymbol{M}$ and two missing data approaches are adopted to include the areas with missing $\boldsymbol{M}$: MICE and EPS imputation. In Scenario 2, $\boldsymbol{M}$ is available directly at the ecological level, whereas scenario 3 mimics the real situation where $\boldsymbol{M}$ is not available directly at the ecological level, but may be estimated through the individual level information.

We find that considering only the survey areas (which can be interpreted as a complete case analysis) provides slightly more biased results and at the same time greater uncertainty than when all the areas are considered and EPS is imputed whenever missing; this is expected as the sample size is smaller ($i \in S$ as opposed to $i \in S \cup I$). In addition using MICE leads to more biased results than using the survey areas only (e.g. bias ranging from 0.04 to 0.10) and is also characterised by a much larger uncertainty (CI95% width ranging from 0.107 to 0.132).

Furthermore, Scenario 3 allows us to assess the impact of individual level sample size (n=5, 10, 20, 100) on the risk estimate for EPS imputation. As expected the bias decreases as the number of sampled individuals increases and the uncertainty tends to decrease slightly; the latter might seem counter-intuitive, but it can be explained by the fact that the number of incomplete areas is kept the same across 3.2.1-3.2.4, to be able to clearly evaluate the gain in accuracy and precision when more individuals are surveyed. In the next section we will show how considering a larger number of surveyed individuals leads to an increase in the proportion of incomplete areas and how this impacts on the $\beta_2$ point estimates and uncertainty.

Table 1 in Supplementary material shows that the same conclusion can be drawn in the absence of an exposure effect ($\beta_2 = 0$).

[Table 1 here]

# 4 Illustrative example: air pollution and health in Greater London

We apply the proposed methodology to investigate the link between air pollution exposure and coronary heart disease (CHD, a subset of cardiovascular diseases) hospital admissions in Greater London. Many studies have provided evidence that short-term exposure (hours or weeks) is associated with cardiovascular disease (CVD) incidence (COMEAP 2006) and large cohort studies in the US (Puett et al. 2009; Miller et al. 2007; Lipsett et al. 2011) and in the Netherlands (Hoek et al. 2002) have focused on the association of long-term exposure to outdoor particulate matter (PM, defined as a mixture of particles smaller than a specific size, e.g. $PM_{10}$ includes particles smaller than 10 micrometers) and CVD related death. However, the epidemiological evidence is more limited regarding the association between long-term exposure to air pollution and CVD morbidity, for instance non-fatal outcomes, such as hospital admissions due to CVD (COMEAP 2006).

We consider the Hospital Episode Statistics (HES), an administrative registry which provides information on admissions at population level and available to us through the Small Area Health Statistics Unit (SAHSU) at Imperial College London. The analysis is conducted for the year 2001 at the electoral ward level, with an average population of about 9,600 persons. To identify the cases, we used the International Classification of Disease version 10 (ICD-10) codes I20 to I25. The number of observed cases in each ward is denoted as $O_i$ (the total number of cases across Greater London was 52,358 in 2001). The total number of residents aged 16 and above in the same region and same year is 5,723,855.

As exposure $X$, we consider the annual mean level of $PM_{10}$ in 2001 at the ward level published by Vienneau et al. (2010) and obtained through a land used regression model combining monitoring data from the national air quality networks (UK) and additional predictors related to traffic, population, land use, and topography. As a binary exposure variable is needed to estimate the EPS as presented in (2), for the purpose of our paper we dichotomise $X$ using the median of the $PM_{10}$ concentration as cut-off ($25\mu g/m^3$).

The confounders $\boldsymbol{C}$ available for every ward at the ecological level are deprivation (measured through the Index of Multiple Deprivation, issued by the Department of Environment Transport and the Regions (2002)) and ethnicity (defined as the proportion of non white and issued by the Census). In addition, the age-sex stratified population count in 2001 census is used to calculate the expected number of hospital admissions $E_i$ in each ward, and thus age and sex are included indirectly in the outcome-exposure analysis.

Following the standard approach in small area studies we first run an ecological regression model which assumes a Poisson distribution on the number of hospitalisations, $O_i \sim$ Poisson$(E_i \lambda_i)$, and a linear regression on $\log \lambda_i$:

$$\log(\lambda_i) = \beta_1 + \beta_2 X_i + \mathbf{C}_i^T \boldsymbol{\beta}_C + \xi_{1i} + \xi_{2i}$$

where $X$ is the exposure, $\mathbf{C}$ are the confounders and $\xi_{1i}$ and $\xi_{2i}$ are spatially unstructured and structured random effects as described in (4). The results show a posterior mean estimate for $\exp(\beta_2)$ (relative risk of CHD hospital admissions for high vs low air pollution exposure) equal to 0.89 (CI95% 0.84 − 0.94). This result is counter-intuitive as it suggests that the risk of hospital admissions is reduced by 11% for the wards with $PM_{10}$ higher than $25 \mu g/m^3$ compared with the wards with the $PM_{10}$ level lower than that threshold. Note that the "protective" effect of air pollution on hospital admissions persists if other categorisations (e.g. quintiles) or a continuous exposure is considered. Residual confounding is the likely cause of this result as the confounders included in the analysis are far from exhaustive given the limited information collected from administrative datasets like HES and Census.

To overcome this issue we integrate additional sources of data providing information on potential confounders via the EPS. We consider the Health Survey for England (HSfE) which is an on-going survey on the health of the individuals living in England and includes each year around 8,000 subjects across the country. Through this we collect information on the following individual level confounders for the years 1994-2001: Education, Smoking, Passive smoking, Drinking, Obesity, Mental illness, Regular exercise, Diabetes, High blood pressure, Vitamin taken, High cholesterol and Table salt intake.

The combined HSfE surveys covered 87.1% of London wards. From the simulation in Section 3.1, it emerged that the number of individual measurements (i.e. sample size $n$ per area) in each area can affect the accuracy of the risk estimates. It might be possible to set a threshold defining the minimum number of subjects per area: above this threshold the areas are included in the EPS calculation, whereas below they are pooled with the incomplete areas (with missing $\boldsymbol{M}$) and then imputed through (3). However this needs to be traded-off against the information lost when areas with some surveyed individuals (albeit a small number) are considered missing. Figure 2(a) shows the histogram of the number of HSfE samples in the wards of London, and Figure 2(b) displays the trade-off between the number of subjects per area and the geographical coverage. If there is at least one subject sampled in any ward, 87.1% of the areas will be used in the EPS estimation, while the remaining 12.9% will be included through the imputation model. Setting the threshold to 5 subjects leads to a drop in coverage to 75.4%, while considering 10 and 20 subjects per ward as threshold lead to 62.6% and 32.0% of the survey coverage, respectively, as shown in Figure 2(b). Due to the low geographical coverage, we did not consider a threshold higher than 20 subjects per ward.

We checked for differences in the values of the measured confounders, outcome and exposure between wards above and below the threshold. For instance, the wards with at least 5 subjects are compared with the ones with less than 5 subjects through density plots in Figure 3, showing the distribution of common shared variables are comparable for these two groups. Based on this we use $\geq 5$ subjects as our main analysis while $\geq 1$, 10 and 20 subjects per ward serve as the sensitivity analysis (See Table 2).

[Figures 2-3 here]

First, we present the results of the analysis on the survey areas, i.e. including only the wards with at least 5 subjects from HSfE ($i \in S$). Table 2 shows that in a standard small area framework, the naïve approach ignoring the confounders $\boldsymbol{M}$ leads to a protective effect of air pollution (posterior mean of the relative risk equal to 0.914 (CI95% $0.89 - 1.00$)). However, after the inclusion of $\boldsymbol{M}$ through EPS adjustment, it changes to 1.08 (CI95% $1.01 - 1.14$). This implies that residual confounding is still substantial in the outcome-exposure relationship in Greater London if only adjusting for the few confounding factors available in all areas (deprivation, ethnicity, age and sex), while the protective effect disappears when including the confounders $\boldsymbol{M}$ as well. By including all the wards in the analysis via EPS imputation, the results are similar with an estimate of 1.03 and a slightly narrower credibility interval (CI95% $0.97 - 1.09$). On the other hand MICE produces a relative risk estimate equal to 0.93 (CI95% $0.88 - 0.99$), suggesting the presence of residual confounding which is not well accounted for by the MICE imputation framework.

Changing the threshold on the minimum number of subjects surveyed in each area does not impact substantially on the results: the point estimates remain stable across the different analyses, which points towards robustness of the estimated relationship between air pollution and CHD. However, when considering only $i \in S$, uncertainty of the estimates increases from 0.13 to 0.21 for the model including the EPS. This is expected as the sample size decreases substantially when the number of surveyed individuals increases. In comparison the model on $i \in S \cup I$ which includes the EPS imputation does not show an important change in uncertainty as the threshold change, which is stable around 0.12-0.14; this suggests that what is gained in precision increasing the number of surveyed individuals is counterbalanced by the increase in uncertainty from the imputation step. In comparison, MICE shows biased results towards the naïve approach, consistently across the different thresholds.

In summary, the result of EPS in Table 2 shows that the wards exposed to a high level of $PM_{10}$ (over 25 $\mu g/m^3$ annually) have on average a 2-4% higher risk of hospital admission than the wards with $PM_{10}$ below that threshold in London, albeit the estimate is characterised by a certain degree of uncertainty. This is somewhat consistent with the individual level study by Cesaroni et al. (2014) which showed a small but positive effect of $PM_{10}$ on coronary events across 11 cohorts in Europe.

[Table 2 here]

## 5 Discussion and Conclusion

In this paper, we developed a novel small area modelling framework based on a propensity score to estimate the effect of an exposure on a health outcome when there are only a limited number of confounders available at the ecological level. We estimate an ecological EPS which synthesises additional individual level confounders from external sources into an area level scalar and define a strategy to impute EPS for the areas where the individual records are missing.

Our EPS is built on the PS defined by McCandless et al. (2012) which considers only the partially measured confounders $\boldsymbol{M}$; this differs from the original definition proposed by Rosenbaum and Rubin (1983) that is $PS_i = \theta_0 + \boldsymbol{C}_i^T \boldsymbol{\theta}_C + \boldsymbol{M}_i^T \boldsymbol{\theta}_M$, which would include also the ecological confounders fully observed on all the areas. Our choice is justified by the fact that, being $\boldsymbol{C}$ fully observed, they should be directly included in the adjustment model. Hence, our interest is to link only $\boldsymbol{M}$ to the outcome-exposure analysis via EPS and at the same time to model the missingness of $\boldsymbol{M}$. An imputation model can be assigned to the partial PS to model the missingness of $\boldsymbol{M}$, but this would not be possible if the original PS definition were used.

We build on from the partial PS and we (i) propose to estimate $\boldsymbol{M}$ from the individual data through a hierarchical model with a spatial structure, (ii) specify a structured imputation model to predict the EPS when missing which accomodates a non linear relationship between the EPS and the $\boldsymbol{C}$ as well as a spatial structure, (iii) block the feedback from the outcome $O$ to the EPS estimation and (iv) allow the feedback from $O$ to the EPS imputation. Points (iii) and (iv) are crucial: in McCandless et al. (2012) the feedback from the outcome influences the estimate of the PS estimation model since this is jointly estimated with the adjustment model. This joint framework has been criticised by Rubin (2007) and recently by Zigler et al. (2013), which in a simulation study showed that the feedback from the outcome leads to a biased estimation of the PS estimates. As we separate the EPS estimate from the analysis model we are not affected by this issue. On the other hand in line with the missing data literature (e.g. Kenward and Carpenter (2007)) the EPS imputation should allow for the input of the outcome $O$ which we ensure through the joint specification of the EPS imputation and of the analysis model.

We run an extensive simulation study to evaluate the performance of our method and to compare it with MICE, a well designed (and commonly used) method to deal with missing data; note that this was the natural model to compare against as a standard Bayesian imputation model would not have been computational viable when considering a relatively large number of confounders, which is the standard case with epidemiological studies. We found that the results from MICE are biased towards the naïve model (i.e. ignoring the confounders $\boldsymbol{M}$) and consider that this might be related to the following two issues: $a$) MICE suffers from the uncongeniality meaning that the functional form of $Y$ in the imputation model is different from its feedback in the analysis model; for instance, in a Poisson model, the outcome has to be included through the $SMR$ to predict EPS in the imputation model, which is different from the feedback function of the observed and expected count in a Poisson likelihood; $b$) in MICE, a missing confounder will be predicted based on other confounders in the same area which are also missing, thus little can be gained from the iterative prediction with the missing confounders as predictors, and this results in increased uncertainty in the estimates; In addition, to ensure a fair comparison we did not include any spatial structure in the simulation study; however it is worth noting that MICE is not designed to take into account the spatial correlation in the imputation stage, while ecological variables are in general spatially correlated, and it is important to account for such correlation in the prediction of missing confounders.

The issues mentioned above can be mitigated by using EPS. First, EPS incorporates the imputation model as a sub-model via a full Bayesian framework, which guarantees the feedback from the outcome to the missing covariate is congenial, thus eliminating the bias arising from imputing a missing covariate in a Poisson likelihood. Second, EPS compresses all missing covariates into one scalar variable, and the single missing EPS can be dealt with more effectively by the imputation model, thus avoiding the issue of using variables with missing data as predictors (as in MICE); as a result, the EPS framework is characterised by less uncertainty than considering the survey areas only. Third, EPS provides the foundation to extend the complexity of the im-putation model via the full Bayesian framework: for instance, spatial correlation can be included as a sub-component of the imputation model. We believe that for a adequate performance of our EPS strategy, the predictive strength of the imputation model in (3) is crucial and should be carefully investigated in each case study. Indeed adequate predictive strength is needed in order to balance the influence of the outcome on the EPS imputation. This is a delicate and understudies issue which will be explored in further work.

By adopting RW(2) as the link function in the PS adjustment, the EPS framework is not only highly flexible in dealing with non-linearity, but can also be implemented to analyse a large ecological dataset due to its much faster speed than the splines based functions. Moreover, RW(2) can be extended to multiple covariates like **C**, and we also considered RW(2) as the ideal function to accommodate the potential non-linearity in the imputation model.

Through the simulation and the application, we showed that EPS is a promising way to incorporate individual level datasets like surveys or cohorts to adjust for unmeasured confounders in small area studies. Future work will extend the framework to deal with categorical instead of binary exposures and to several correlated exposure variables which are particularly relevant aspects in the field of environmental epidemiology aiming at feeding back policy makers on the epidemiological risk of environmental hazards.

# 6 Ackowledgments

The work of the UK Small Area Health Statistics Unit is funded by Public Health England as part of the MRC-PHE Centre for Environment and Health, funded also by the UK Medical Research Council. SAHSU holds approvals from the National Research Ethics Service - reference 12/LO/0566 and 12/LO/0567 - and from the Health Research Authority Confidentially Advisory Group (HRA-CAG) for Section 251 support (HRA - 14/CAG/1039). Marta Blangiardo, Anna Hansell, Sylvia Richardson and Monica Pirani acknowledge support from the MRC Methodology grant MR/M025195/1. Sylvia Richardson acknowledges support from the MRC programme grant MC_UP_0801/1. The authors thank Nicola Best for fruitful discussion on the developed methodology, Gianluca Baio and Alexina Mason for providing comments on the draft of the manuscript.

# Figures and tables

Figure 1: The figure represents the developed EPS framework. The left hand side corresponds to (1) and (2): the area level confounders $M$ are estimated from the individual level confounders $m$ (1) and the EPS is estimated (2). Note that this model is only specified on the $i \in S$ areas. EPS is represented by a circle as a latent quantity (not observed). The right hand side presents the EPS imputation and adjustment and is specified on $i \in I \cup S$: at the top the EPS is either obtained from (2) for $i \in S$ and included as an observed quantity or missing for $i \in I$, imputed through the relationship with $C$, $X$ and the spatial structure (3) and included as a latent quantity. Thus the EPS is represented here as both a square and a circle as the combination of observed and latent quantities. At the bottom, the estimated and imputed EPS is included in the analysis model (corresponding to (4)) to provide confounder adjustment when assessing the effect of $X$ on $O$. Note that the right hand side is separated from the left hand side as the EPS estimation is carried out disjointly from the imputation and adjustment (no feedback from the latter to the former). At the same time the imputation and adjustment are jointly carried out so that feedback is allowed from $O$ to influence the EPS imputation.

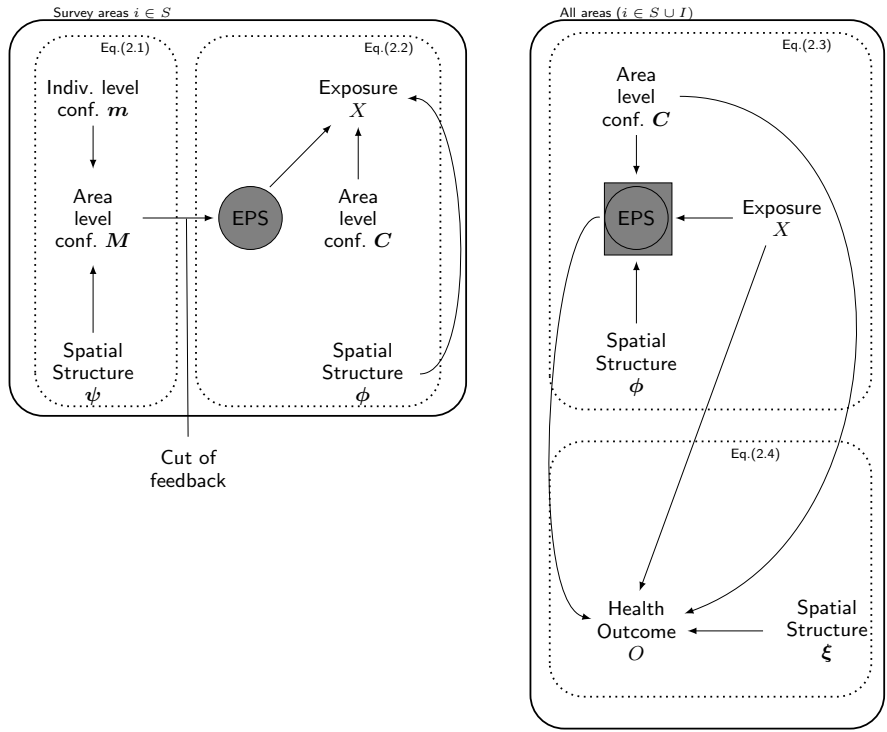

Figure 2: HSfE sampled individuals and geographical coverage. The left hand side shows the distribution of the number of individuals sampled through HSfE across the wards; the right hand side shows the trade-off between spatial coverage and the number of sampled individuals. The four dashed lines represent the thresholds considered for estimating / imputing EPS (1, 5, 10 and 20 subjects per ward).

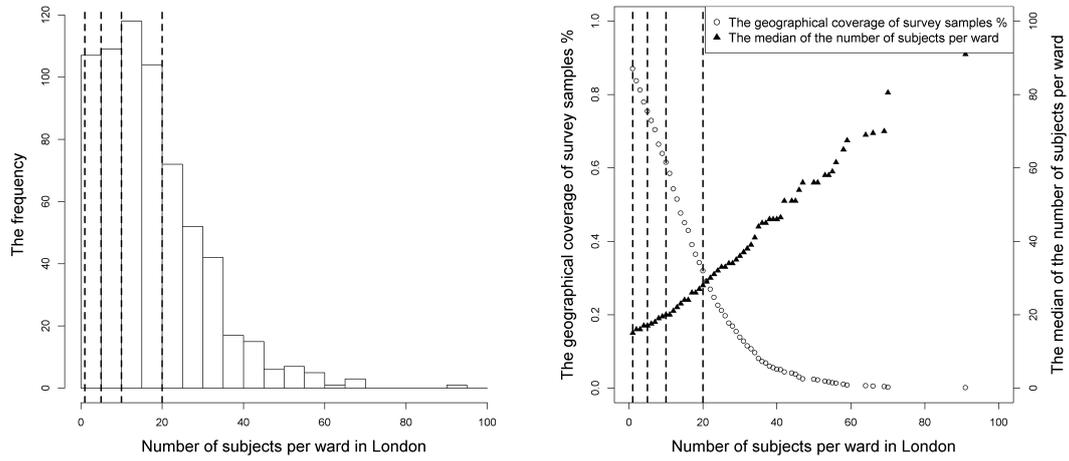

Figure 3: Comparison of wards covered/not covered by HSfE. The four density plots represent the distribution of the following variables across wards (from left to right): (i) SMR=$O/E$; (ii) $PM_{10}$, included as a continuous variable ($\mu g/m^3$); (iii) IMD01, the Index of Multiple Deprivation at 2001; (iv) Proportion of non-white individuals. The solid lines are the distributions in the wards with $\geq 5$ individuals sampled by HSfE, while the dashed line the distributions for the wards with $\leq 5$ individuals sampled by HSfE or not covered by HSfE. A similar behaviour can be seen for all the variables across wards.

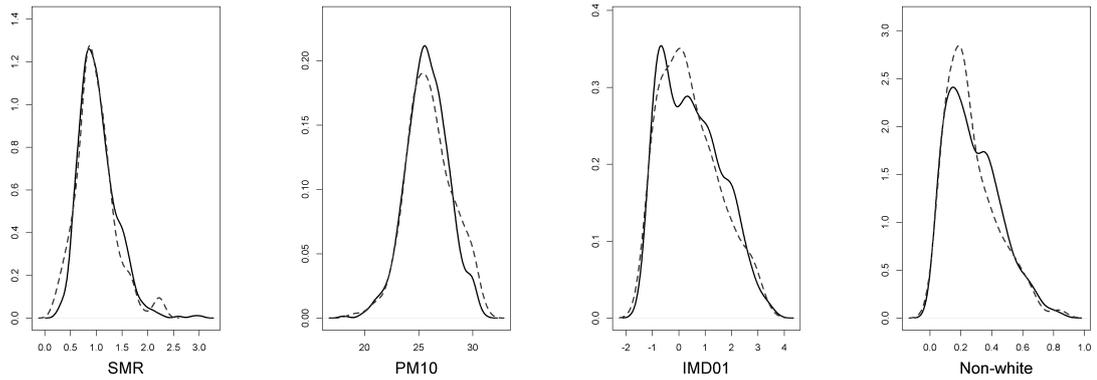

Table 1: EPS performance and comparison with MICE on the simulation study (true $\beta_2 = 0.5$, 100 simulated datasets).

|  | Adjustment/ Imputation | Posterior Mean for $\beta_2$ | Bias | Root Mean Square Error | CI95% width |
|---|---|---|---|---|---|
| **Scenario 1: $M$ are available in all areas** | | | | | |
|  | Direct adj | 0.50 | 0.00 | 0.02 | 0.054 |
|  | EPS adj | 0.50 | 0.00 | 0.02 | 0.062 |
| **Naïve case: Ignoring $M$** | | | | | |
|  | NA | 0.78 | 0.28 | 0.28 | 0.049 |
| **Scenario 2: $M$ are only available in some areas** | | | | | |
| Case 2.1: Analysis on $i \in S$ | EPS adj | 0.51 | 0.01 | 0.03 | 0.087 |
| Case 2.2 : Analysis on $i \in S \cup I$ | MICE | 0.54 | 0.04 | 0.05 | 0.107 |
|  | EPS imput | 0.50 | 0.00 | 0.03 | 0.084 |
| **Scenario 3: $M$ are NOT directly available, but $m$ are available in some areas** | | | | | |
| **Sample size n=5** | | | | | |
| Case 3.1.1: Analysis on $i \in S$ | EPS adj | 0.59 | 0.08 | 0.09 | 0.094 |
| Case 3.2.1: Analysis on $i \in S \cup I$ | MICE | 0.62 | 0.12 | 0.14 | 0.132 |
|  | EPS imput | 0.57 | 0.07 | 0.08 | 0.087 |
| **Sample size n=10** | | | | | |
| Case 3.1.2: Analysis on $i \in S$ | EPS adj | 0.59 | 0.09 | 0.10 | 0.094 |
| Case 3.2.2: Analysis on $i \in S \cup I$ | MICE | 0.62 | 0.12 | 0.14 | 0.129 |
|  | EPS imput | 0.57 | 0.07 | 0.08 | 0.086 |
| **Sample size n=20** | | | | | |
| Case 3.1.3: Analysis on $i \in S$ | EPS adj | 0.59 | 0.09 | 0.10 | 0.094 |
| Case 3.2.3: Analysis on $i \in S \cup I$ | MICE | 0.60 | 0.10 | 0.11 | 0.128 |
|  | EPS imput | 0.57 | 0.07 | 0.08 | 0.085 |
| **Sample size n=100** | | | | | |
| Case 3.1.4: Analysis on $i \in S$ | EPS adj | 0.52 | 0.02 | 0.04 | 0.088 |
| Case 3.2.4: Analysis on $i \in S \cup I$ | MICE | 0.56 | 0.06 | 0.06 | 0.119 |
|  | EPS imput | 0.51 | 0.01 | 0.03 | 0.083 |

Table 2: Relative risk of hospital admission: $PM_{10}$ higher than $25\mu g/m^3$ (chosen cut point) in London. The main analysis with $\geq 5$ subjects per wards is presented in bold, while the others are used as sensitivity analyses.

| Areas | Data used | $\geq 1$ subject (12.9% missing) | | $\geq$ **5 subjects** **(24.6% missing)** | | $\geq 10$ subjects (37.4% missing) | | $\geq 20$ subjects (68.0% missing) | |
|---|---|---|---|---|---|---|---|---|---|
| | | RR (CI95%) | CI95% width | RR (CI95%) | CI95% width | RR (CI95%) | CI95% width | RR (CI9%5) | CI95% width |
| $i \in S$ | X,**C** | 0.91 (0.86-0.96) | 0.10 | **0.94 (0.89-1.00)** | **0.11** | 0.96 (0.90-1.03) | 0.13 | 0.96 (0.88-1.04) | 0.16 |
| | X,**C**,EPS | 1.05 (0.98-1.11) | 0.13 | **1.08 (1.01-1.14)** | **0.13** | 1.09 (1.01-1.18) | 0.17 | 1.06 (0.96-1.17) | 0.21 |
| $i \in S \cup I$ | X,**C** | 0.89 (0.84-0.94) | 0.10 | **0.89 (0.84-0.94)** | **0.10** | 0.89 (0.84-0.94) | 0.10 | 0.89 (0.84-0.94) | 0.10 |
| | X,**C**,EPS | 1.04 (0.98-1.10) | 0.12 | **1.03 (0.97-1.09)** | **0.12** | 1.05 (0.98-1.12) | 0.14 | 1.05 (0.99-1.12) | 0.13 |
| | X,**C**,**M** (MICE) | 0.98 (0.93-1.04) | 0.11 | **0.93 (0.88-0.99)** | **0.11** | 0.92 (0.87-0.98) | 0.11 | 0.89 (0.84-0.94) | 0.10 |

# Using Ecological Propensity Score to Adjust for Missing Confounders in Small Area Studies

Yingbo Wang

*Novartis, Basel (Switzerland)*

Sylvia Richardson

*MRC Biostatistics Unit, Cambridge University (UK)*

Monica Pirani, Anna Hansell, Marta Blangiardo∗

*MRC-PHE Centre for Environment and Health, Imperial College London (UK)*

m.blangiardo@imperial.ac.uk

∗ To whom correspondence should be addressed.

## APPENDIX

### A. Multivariate CAR

The BYM model proposed by Besag et al. (1991) can be assigned to $\psi_{1qd} + \psi_{2qd}$ on district $d$:

$$\psi_{1qd}|(\psi_{1ql,\ l\neq d}) \sim N\left(\frac{\sum_{l\in\mathcal{N}(d)} \psi_{1ql}}{\mathcal{N}(d)}, \frac{\sigma^2_{\psi_{1q}}}{\mathcal{N}(d)}\right) \quad (A.1)$$

$$\psi_{2qd} \sim N(0, \sigma^2_{\psi_{2q}})$$

$$\alpha_q \sim \text{Uniform}(-\infty, +\infty)$$

where $\psi_{1qd}$ is specified through the intrinsic conditional autoregressive model (iCAR) proposed by Besag and Kooperberg (1995), while $\psi_{2qd}$ follows a normal distribution with a common variance $\sigma^2_{\psi_{2q}}$. For the $q$th confounder, a local smoothing is provided by $\psi_{1qd}$ based on the values of the set of neighbours $\mathcal{N}(d)$ and a global smoothing is included through $\psi_{2qd}$ based on the values of all the units. The intercept $\alpha_q$ needs to be drawn from a flat distribution between $\pm\infty$ in order to make iCAR identifiable.

A structure of the correlation between the confounders should be included to allow borrowing of strength, as some are observed on more individuals than others (e.g. some confounders might have been collected for several years while others might not). Thus, we extend $\sigma^2_{\psi_{1q}}$ in (A.1) further to $\Sigma^2_{\psi_1}$ (the diagonal of $\Sigma^2_{\psi_1}$ is $\sigma^2_{\psi_{1q}, \, q=1,\ldots,Q}$ and the off diagonal is $\rho_{\psi'_{1qq}}\sigma_{\psi_{1q}}\sigma_{\psi_{1q'}}, \; q'\neq q$), and similarly the between confounder non-spatial correlation $\rho_{\psi'_{2qq}}$ is taken into account by replacing $\sigma^2_{\psi_{2q}}$ with $\Sigma^2_{\psi_2}$, becoming the multivariate BYM model (MVBYM). See Gamerman et al. (2003), Thomas et al. (2004) for details and applications of MVBYM.

## B. The simulation process

The data simulation process is described below:

1. Simulate one set of ecological variables $X, \mathbf{C} = (C_1, C_2), \mathbf{M} = (\mathbf{M}_1, \mathbf{M}_2, \mathbf{M}_3, \mathbf{M}_4)$. Let $i$ denote the area index with $i = 1, \ldots, 300$ ($i \in S \cup I$). Simulate $\mathbf{C}$ based on the expit transformation of $N(0,1)$, and generate correlated $\mathbf{M}$ based on the expit transformation of bivariate normal distribution:

$$\begin{pmatrix} \text{logit}(M_{1i}) \\ \text{logit}(M_{2i}) \end{pmatrix} \sim \text{MVN}_2 \left( \begin{bmatrix} 0 \\ 0 \end{bmatrix}, \sigma^2 \begin{bmatrix} 1 & 0.3 \\ 0.3 & 1 \end{bmatrix} \right)$$

$$\begin{pmatrix} \text{logit}(M_{3i}) \\ \text{logit}(M_{4i}) \end{pmatrix} \sim \text{MVN}_2 \left( \begin{bmatrix} 0 \\ 0 \end{bmatrix}, \sigma^2 \begin{bmatrix} 1 & 0.3 \\ 0.3 & 1 \end{bmatrix} \right)$$

where $\sigma = 1$. Then the exposure $X$ is produced through a Bernoulli distribution:

$$X_i \sim \text{Bernoulli}(P(X_i = 1 | \mathbf{C}_i, \mathbf{M}_i))$$

$$\text{logit}(P(X_i = 1 | \mathbf{C}_i, \mathbf{M}_i)) = \theta_1 + \mathbf{C}_i^T \boldsymbol{\theta}_C + \mathbf{M}_i^T \boldsymbol{\theta}_M$$

$$\text{EPS}_i = \mathbf{M}_i^T \boldsymbol{\theta}_M$$

Ecological Propensity Score for Small Area Studies                                   3where $\boldsymbol{\theta}_C = (\theta_2, \theta_3)$ and $\boldsymbol{\theta}_M = (\theta_4, \theta_5, \theta_6, \theta_7)$. The true values for $\boldsymbol{\theta}$ are set to be the following:

$$\theta_1 = 0; \theta_2 = 0.5; \theta_3 = -0.5; \theta_4 = 1; \theta_5 = -0.6; \theta_6 = 0.5; \theta_7 = -0.4$$

Suppose the expected count $E$ is the same across all areas, i.e. $E_i = 100 \; \forall \; i = 1, \ldots, 300$, then the observed count $O$ is simulated through:

$$O_i \sim \text{Poisson}(E_i \lambda_i)$$

$$\log(\lambda_i) = \beta_1 + \beta_2 X_i + \mathbf{C}_i^T \boldsymbol{\beta}_C + \mathbf{M}_i^T \boldsymbol{\beta}_\pi \tag{B.1}$$

The true values for $\beta_1, \beta_2, \boldsymbol{\beta}_C = (\beta_3, \beta_4)$ and $\boldsymbol{\beta}_M = (\beta_5, \beta_6, \beta_7, \beta_8)$ are:

$$\beta_1 = 0; \beta_2 = 0.5 \text{ or } 0; \beta_3 = 0.2; \beta_4 = -0.2; \beta_5 = 0.2; \beta_6 = -0.2; \beta_7 = 0.2; \beta_8 = -0.2$$

2. Simulate $n$ values for the individual variables $\boldsymbol{m} = (\boldsymbol{m}_1, \boldsymbol{m}_2, \boldsymbol{m}_3, \boldsymbol{m}_4)$:

$$m_{qij} \sim \text{Bernoulli}(M_{qi}), q = 1, ..., 4, \; j = 1, ..., n$$

where $n$ is chosen to be 5, 10, 20 and 100 for the different simulation scenarios.

3. The following MAR criterion is then applied to remove $\boldsymbol{M}$ or $\boldsymbol{m}$ from around 50% of the areas:

$$l_i = \text{Bernoulli}(P(l_i = 1)) \tag{B.2}$$

$$\text{logit}(P(l_i = 1)) = 0.2 C_{1.i} - 0.2 C_{2.i}$$

where $l$ is the indicator for the missingness of $\boldsymbol{M}$ or $\boldsymbol{m}$. The complete cases are defined as the areas with $\boldsymbol{M}_i$ or $\boldsymbol{m}_i$ available ($l_i = 0$, i.e. $i \in S$), while the remaining areas have missing $\boldsymbol{M}_i$ or $\boldsymbol{m}$, i.e. $i \in I$.

The simulation result with true $\beta_2 = 0$ is shown in Table 1.

Table 1: EPS performance and comparison with MICE on the simulation study (true $\beta_2 = 0$, 100 simulated datasets).

| | Adjustment/ Imputation | Posterior Mean for $\beta_2$ | Bias | Root Mean Square Error | CI95 width |
|---|---|---|---|---|---|
| **Scenario 1: $M$ are available in all areas** | | | | | |
| | Direct adj | 0.00 | 0.00 | 0.02 | 0.060 |
| | EPS adj | 0.00 | 0.00 | 0.02 | 0.064 |
| **Naïve case: Ignoring $M$** | | | | | |
| | NA | 0.28 | 0.28 | 0.28 | 0.054 |
| **Scenario 2: $M$ are only available in some areas** | | | | | |
| Case 2.1: Analysis on $i \in S$ | EPS adj | 0.00 | 0.00 | 0.03 | 0.093 |
| Case 2.2 : Analysis on $i \in S \cup I$ | MICE | 0.03 | 0.03 | 0.04 | 0.081 |
| | EPS imput | -0.01 | -0.01 | 0.03 | 0.088 |
| **Scenario 3: $M$ are NOT directly available, but $m$ are available in some areas** | | | | | |
| **Sample size n=5** | | | | | |
| Case 3.1.1: Analysis on $i \in S$ | EPS adj | 0.09 | 0.09 | 0.11 | 0.087 |
| Case 3.2.1: Analysis on $i \in S \cup I$ | MICE | 0.13 | 0.13 | 0.14 | 0.142 |
| | EPS imput | 0.08 | 0.08 | 0.08 | 0.092 |
| **Sample size n=10** | | | | | |
| Case 3.1.2: Analysis on $i \in S$ | EPS adj | 0.09 | 0.09 | 0.11 | 0.088 |
| Case 3.2.2: Analysis on $i \in S \cup I$ | MICE | 0.13 | 0.13 | 0.13 | 0.139 |
| | EPS imput | 0.08 | 0.08 | 0.08 | 0.091 |
| **Sample size n=20** | | | | | |
| Case 3.1.3: Analysis on $i \in S$ | EPS adj | 0.08 | 0.08 | 0.10 | 0.085 |
| Case 3.2.3: Analysis on $i \in S \cup I$ | MICE | 0.10 | 0.10 | 0.11 | 0.139 |
| | EPS imput | 0.07 | 0.07 | 0.08 | 0.089 |
| **Sample size n=100** | | | | | |
| Case 3.1.4: Analysis on $i \in S$ | EPS adj | 0.02 | 0.02 | 0.05 | 0.091 |
| Case 3.2.4: Analysis on $i \in S \cup I$ | MICE | 0.06 | 0.06 | 0.06 | 0.126 |
| | EPS imput | 0.01 | 0.01 | 0.04 | 0.089 |